\documentclass[%
reprint,
physrev]{revtex4-2}

\usepackage{amsmath} 
\usepackage{graphicx}
\usepackage{dcolumn}
\usepackage{bm}

\usepackage{hyperref}
\hypersetup{
    colorlinks=true,
    citecolor = blue
}

\usepackage[abs]{overpic}
\newcommand{\p}{\partial}

\newcommand{\la}{\langle}
\newcommand{\ra}{\rangle}

\begin{document}

\title{\textbf{Inviscid scaling in the Kuramoto-Sivashinsky equation from functional renormalization group and direct numerical simulations}
}%

\author{Liubov Gosteva$^1$, Dipankar Roy$^2$, Nicol\'as Wschebor$^3$, L\'eonie Canet$^1$}
\email{Corresponding author: leonie.canet@lpmmc.cnrs.fr}
\affiliation{$^1$Univ. Grenoble Alpes, CNRS, LPMMC, 38000 Grenoble, France\\ $^2$ School of Physics, Indian Institute of Science Education and Research Thiruvananthapuram, Maruthamala PO, Thiruvananthapuram, Kerala 695551, India\\ $^3$Instituto de F\'isica, Facultad de Ingenier\'ia, Universidad de la Rep\'ublica, J.H.y Reissig 565, 11000 Montevideo, Uruguay}

\begin{abstract}
We show that the one-dimensional Kuramoto-Sivashinsky (KS) equation features a scaling regime characterized by the dynamical exponent $z=1$ at intermediate scales between the large-scale Kardar-Parisi-Zhang (KPZ) scaling with $z=3/2$ and the small-scale non-universal behavior. This scaling regime is intrinsic to the KS dynamics since it arises from the vanishing of the effective viscosity when evolving from its microscopic negative KS value, to its macroscopic effective positive KPZ value. This vanishing of the viscosity deeply imprints the behavior of correlations at  intermediate scales, which exhibit a universal $z=1$ scaling. This  behavior  pertains to the inviscid-Burgers universality class, which corresponds to the  zero-viscosity fixed point of the KPZ equation.  We evidence and characterize this so-far-overlooked scaling regime using both functional renormalization group and direct numerical simulations.
\end{abstract}

\maketitle

The Kuramoto-Sivashinsky (KS) equation describes the dynamics of a one-dimensional
scalar field $h$ as
\begin{equation} 
\p_t h(t,x) = \nu \p_x^2 h(t,x) - \tau \p_x^4 h(t,x) + \frac{\lambda}{2} (\p_x h(t,x))^2
\label{eq:KS}
\end{equation}
where the viscosity $\nu$ is negative $\nu<0$, and $\tau$ is positive $\tau>0$ to ensure dynamical stability. 
 It models a great variety of physical phenomena exhibiting chaos and instabilities: chemical turbulence in reaction-diffusion systems \cite{KuramotoTsuzuki1975, KuramotoTsuzuki1976, Kuramoto1978}; flame front instabilities \cite{Sivashinsky1977_1, MichelsonSivashinsky1977_2, Sivashinsky1977, Sivashinsky1980};
flow of a liquid down an inclined plane  \cite{Nepomnyashchii1974}, vertical plane   \cite{SivashinskyMichelson1980} or pipe \cite{Chen1986}; 
plasma instabilities associated with trapped ions \cite{LaQuey1975};
surface roughening induced by heavy ion irradiation \cite{Cuerno1995, Kanjilal2006}; dendritic patterns evolution during alloy solidification \cite{Losert1998}.
 The KS equation also emerges in the context of the complex Ginzburg-Landau equation (CGLE) as the effective dynamics of the phase of the complex order parameter, when its amplitude weakly fluctuates (phase turbulence regime) \cite{Kuramoto1978,Grinstein1996,Aranson2002}.

The large-scale behavior of the KS equation belongs to the celebrated Kardar-Parisi-Zhang (KPZ) universality class, which means that the effective dynamics of the coarse-grained  field is well described at large distances and long times by the KPZ equation \cite{Kardar1986}
\begin{equation}
\p_t h = \nu_{\rm eff} \p_x^2 h + \frac{\lambda}{2} (\p_x h)^2 + \eta\,, 
\label{eq:KPZ}
\end{equation}
with a positive effective viscosity $\nu_{\rm eff}>0$ and an effective Gaussian noise 
  $\eta(t,x)$ of zero mean and covariance
\begin{equation}
\la\eta(t,x)\eta(t',x')\ra = 2 D_{\rm eff}\delta(t-t')\delta(x-x')\,.
\label{eq:etaKPZ}
\end{equation}
This was suggested within Mori’s method \cite{Fujisaka1977},  using renormalization group (RG) approach \cite{Yakhot1980, Ueno2005}, and by construction of an effective large-scale model \cite{Zaleski1989}. The mechanism is that the intrinsic instabilities of the KS equation leads to a chaotic dynamics, which in turn generates an effective stochastic noise and a positive effective viscosity. However, the actual confirmation of the KPZ scaling $z=3/2$, where $z$ is the dynamical critical exponent, in numerical simulations of the {\it deterministic} KS equation was provided only recently \cite{Roy2020}  since it requires huge system size and simulation time. Early  numerical studies \cite{Hyman1986,Zaleski1989, Sneppen1992, Hayot1993, Ueno2005}  reported instead the diffusive scaling $z=2$ pertaining to the linear ($\lambda=0$) Edwards-Wilkinson (EW) equation. For the KPZ equation, the persistence of EW scaling is observed when the system size is small compared to $1/g_{\rm KPZ}$ where $g_{\rm KPZ}=\lambda^2 D_{\rm eff}/\nu_{\rm eff}^3$.
For the deterministic KS equation, the corresponding coupling $g_{\rm KS}$ is vanishing since the noise amplitude is zero, and it only builds up gradually. On the other hand, the finite-size effect can be overcome by introducing a noise into the KS equation, which henceby becomes {\it stochastic}, and exhibits KPZ scaling for much smaller system sizes and times \cite{Ueno2005}.

In this paper, we demonstrate the existence of another scaling regime, characterized by $z=1$, which arises over an extended range of wavenumbers intermediate between the small ones (large distance) featuring KPZ scaling, and the largest unstable mode. This scaling regime is very robust and is observed irrespective of the large-distance scaling (KPZ $z=3/2$ or EW $z=2$ one).
It intrinsically originates from the KS dynamics, and can be understood from the renormalization group coarse-graining process. When integrating rapid (high-momentum) modes with wavenumbers larger than the RG scale $\kappa$, the infrared (IR) modes develop an effective scale-dependent viscosity $\nu_\kappa$ evolving from the KS microscopic value $\nu<0$ to a value at large distances (when $\kappa\to 0$)  $\nu_{\rm eff}\equiv \nu_{\kappa=0}$ which is positive as in the KPZ  equation.  Thus, correlation functions for wavenumbers of the order of the scale $\kappa$ for which $\nu_{\kappa} \simeq 0$ are approximately described by an effective KPZ equation with vanishing viscosity.
 This corresponds to a specific, unstable fixed point of the KPZ equation, which was recently unveiled and called the Inviscid Burgers (IB) fixed point \cite{Fontaine2023Unpredicted, Gosteva2025Burgers, Gosteva2025KPZtwogrids}. It differs from the KPZ fixed point obtained at finite $\nu$ which yields the well-known KPZ scaling $z=3/2$. Instead, the IB fixed point is characterized by $z=1$ and its own non-trivial scaling function  \cite{Fontaine2023Unpredicted, Gosteva2025Burgers}.  In the KS equation, the IB scaling generically  arises, without requiring fine-tuning of parameters (such as choosing a small enough $\nu$ in the KPZ equation), and it turns out to control an extended  range of  intermediate scales. We show this using both a functional renormalization group (FRG) analysis and direct numerical simulations (DNS) of the KS equation.

Let us mention that an observation of the IB scaling was reported in simulations of the complex Ginzburg-Landau equation in the phase turbulence regime,  where the KS equation emerges as an effective description  of the dynamics of the phase  of the complex order parameter \cite{Vercesi2024}. Here, we provide evidence for the KS equation itself, and both from simulations and from FRG. We first briefly present these two approaches before discussing the results.

\noindent{\it FRG analysis.} The  different scaling regimes can be identified from the two-point space-time correlation function $C(t,x) = \langle h(t,x)  h(0,0) \rangle_c$ in the stationary state, where $c$ stands for connected. The calculation of the full space-time, or equivalently frequency-momentum correlation function requires a functional approach. Moreover, the IB fixed point is genuinely non-perturbative as it corresponds to $g_{\rm KPZ}\to \infty$.  This justifies the use of FRG.

The FRG formalism \cite{Wetterich1993, Delamotte2012, Dupuis2021} is a functional implementation of Wilson's RG. It is formulated on the path integral representation obtained from the Martin-Siggia-Rose-Janssen-de Dominicis procedure \cite{Martin1973,Janssen1976,Dominicis1976}. This procedure is built on the basis of a stochastic equation. We therefore introduce a noise, of the form \eqref{eq:etaKPZ} with amplitude $D$, in the KS equation, as done in \cite{Yakhot1980, Ueno2005}, whose purpose it to replace the average over initial conditions in the deterministic equation by an average over noise realizations. This noise modifies the scale at which the  IR KPZ scaling emerges, but it does not affect intermediate scales, as was demonstrated in \cite{Vercesi2024}. The limit $D\to0$ will be studied elsewhere. The KS action is given by
\begin{equation}
{\cal S}_{\Lambda}[h,\bar h] = \int_{t,x} \left[ \bar h(t,x) E[h(t,x)] - D\bar h(t,x)^2 \right] \,,
\label{eq:Sbare}
\end{equation}
where $E[h]= \p_t h - \nu \p_x^2 h + \tau \p_x^4 h - (\lambda/2) (\p_x h)^2$ is the deterministic KS equation, $\bar h$ is the response field. The index  $\Lambda$  denotes the microscopic (UV) momentum scale at which the KS equation is defined (which can be chosen very large).
The FRG method consists in progressively averaging over fluctuation modes  by introducing a scale-dependent weight
\begin{align}
\Delta \mathcal{S}_\kappa = \int_{t,x,x'}&\Big\{-\bar{h}(t,x) R^\nu_\kappa(x-x') \nabla'^2 h(t,x')\nonumber\\
& - \bar{h}(t,x) R^D_\kappa(x-x') \bar{h}(t,x') \Big\}\,,
\label{eq:deltaSk}
\end{align}
 where $\kappa$ is the RG momentum scale. The $\kappa$-dependent generating functional is
$\mathcal{Z}_\kappa[j,\bar{j}] = \int \mathcal{D} h \mathcal{D} \bar{h}\, \exp\left({-{\mathcal{S}_\kappa}[h,\bar{h}] + \int_{t,x}(j h + \bar j \bar{h})} \right) $ 
 where  $j,\bar j$ are the sources, and
$\mathcal{S}_\kappa = \mathcal{S}_\Lambda + \Delta \mathcal{S}_\kappa$.
One then defines the effective average action $\Gamma_{\kappa}[\la h \ra , \la \bar{h} \ra]= -\ln\mathcal{Z}_\kappa + \int_{t,x} \big(j \la h \ra + \bar{j} \la \bar{h} \ra\big) - \Delta \mathcal{S}_\kappa[\la h \ra , \la \bar{h} \ra]$. Its evolution with the RG scale is given by   Wetterich equation \cite{Wetterich1993}
\begin{equation}  
    \partial_{\kappa} \Gamma_{\kappa} =
    \frac{1}{2}\textrm{tr}\,\int_{\omega,q}
    \partial_{\kappa} \mathcal{R}_{\kappa}\,
    \mathcal{G}_{\kappa}\,,\;\;\;\; \mathcal{G}_{\kappa} \equiv \left(
    \Gamma_{\kappa}^{(2)} + \mathcal{R}_{\kappa}
\right)^{-1},
\label{eq:Wetterich}
\end{equation}
where the trace means summation over all fields $\la h \ra$,$\la \bar h\ra$,
$\mathcal{G}_{\kappa}$  is the propagator matrix, $\Gamma_{\kappa}^{(2)}$ is the Hessian of  $\Gamma_\kappa$, and
$\mathcal{R}_{\kappa}$  is the regulator matrix 
 defined as the Hessian of  $\Delta \mathcal{S}_\kappa$.

The regulators $R^{\nu,D}_\kappa(x)$ in \eqref{eq:deltaSk} can be chosen arbitrarily provided they have the following properties in Fourier space: they are large $R^{\nu,D}_\kappa(p)\sim \kappa^2$ at small momenta $p\lesssim\kappa$ and vanish $R^{\nu,D}_\kappa(p)\simeq 0$ at large $p$. The regulator term plays the role of a large mass for slow modes ($p\lesssim \kappa$) such that they do not contribute to the functional integration in ${\cal Z_\kappa}$, while they leave the fast modes ($p\gtrsim \kappa$)  unaffected such that they are integrated in. The regulators therefore achieve the progressive averaging over the fluctuation modes, and the effective average action $\Gamma_\kappa$ smoothly interpolates between the bare microscopic action (\ref{eq:Sbare}) at $\kappa=\Lambda$  and the effective large-lengthscale description at $\kappa=0$ which contains all the statistical properties of the system.

The FRG equation (\ref{eq:Wetterich}) is exact, but in general can be solved only approximately, using well-established approximation schemes \cite{Dupuis2021}. In this work we use the NLO (Next-to-Leading-Order) approximation, developed for the KPZ equation in  Ref.~\cite{Kloss2012}.
In a nutshell, the coefficients $\nu$ and $D$ are promoted to $\kappa$-dependent functions $f^\nu_\kappa(\omega,p)$ and $f^D_\kappa(\omega,p)$ of frequency $\omega$ and momentum $p$. One defines the effective noise amplitude as $D_\kappa\equiv f^D_\kappa(0,0)$ and effective viscosity as $\nu_\kappa\equiv f^\nu_\kappa(0,0)$.
To study scaling regimes, one introduces dimensionless functions  $\hat f^\nu_\kappa(\hat\omega,\hat p) = f^\nu_\kappa(\omega,p)/A_\kappa$, $\hat f^D_\kappa(\hat\omega,\hat p) = f^D_\kappa(\omega,p)/D_\kappa$, where $\hat\omega=\omega/(\kappa^2\nu_\kappa)$ and $\hat p=p/\kappa$.  For the KPZ equation, the usual choice is $A_\kappa=\nu_\kappa$, but it is not appropriate for the KS equation since the latter is expected to cross zero  during the renormalization flow. We use instead $A_\kappa \equiv |\nu_\Lambda| \kappa^{-\eta_\nu}$, where we set $\eta_\nu=1/2$ the exact  exponent expected at the KPZ fixed point. 
The RG flow equations for the functions $f^{\nu,D}_\kappa$ are obtained from the Wetterich equation \eqref{eq:Wetterich} and can be found in Refs.~\cite{Kloss2012,Gosteva2025KPZtwogrids}.
These equations determine the evolution of the functions under the change of scale $\kappa$. We  start from the KS initial conditions 
$\nu_\Lambda = \nu \equiv -1$, $\tau_\Lambda = \tau \equiv 1$, $D_\Lambda=D \equiv 1$,
$\hat f^D_\Lambda = 1$, $\hat f^\nu_\Lambda = -1 + \hat p^2$ (we set $\Lambda=1$).  
We  solve the dimensionless flow evolution, until we reach when $\kappa\to 0$ a fixed point, which will be identified in the following as the KPZ one.

\begin{figure}
    \begin{overpic}[width=9cm]{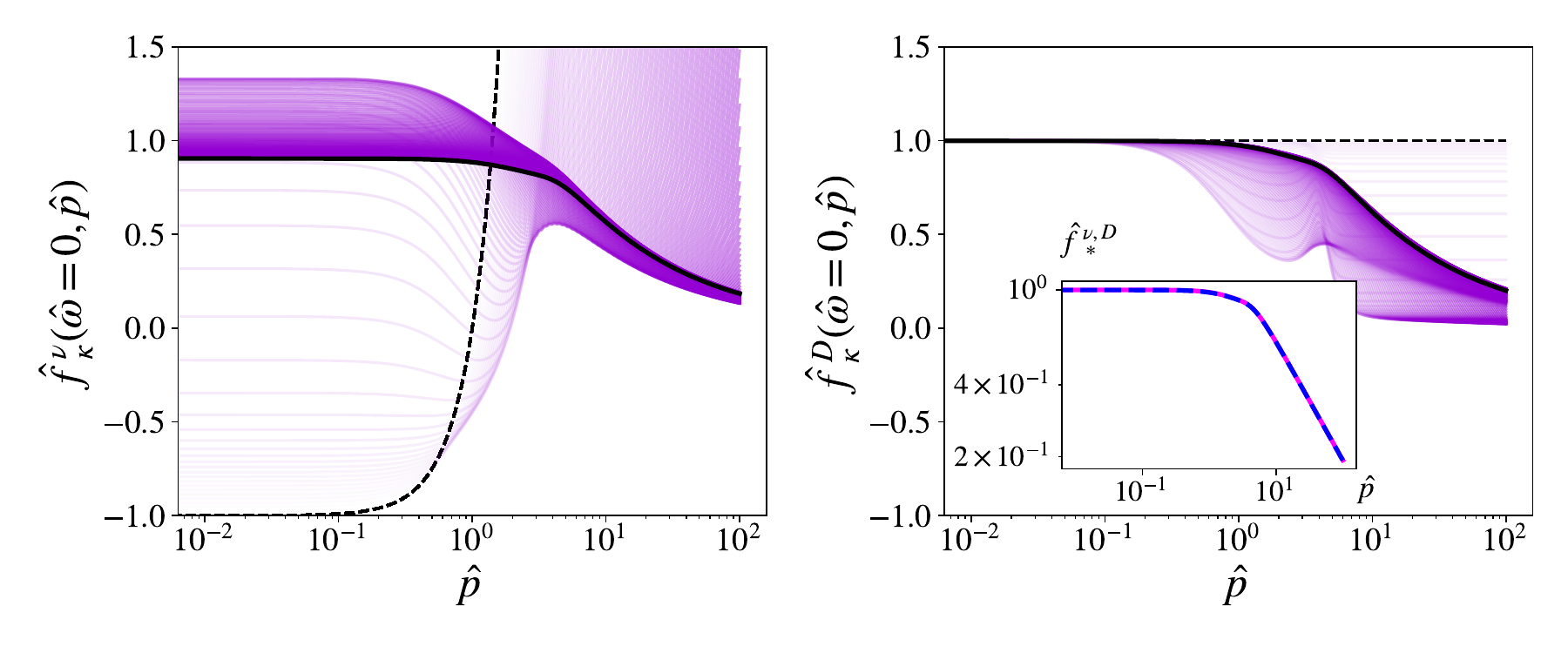}
        \put(7,90){\scriptsize(a)}
        \put(135,90){\scriptsize(b)}
    \end{overpic}
    \caption{Evolution of the functions (a) $\hat f^\nu_\kappa$ and (b) $\hat f^D_\kappa$ under the FRG flow, starting from the KS initial condition $\hat f^D_\Lambda = 1$, $\hat f^\nu_\Lambda = -1 + \hat p^2$  (black dashed lines) 
    and reaching a fixed point when $\kappa\to 0$ (black solid lines correspond to $\kappa/\Lambda=e^{-8}$). Inset of (b): Fixed point shape of $\hat f^\nu_\kappa(\hat \omega=0,\hat p)$ (blue, dashed) and $\hat f^D_\kappa(\hat \omega=0,\hat p)$ (magenta, plain) recorded at $\kappa/\Lambda=e^{-400}$. They are identical, which indicates that time-reversal symmetry emerges at large distances, and they identify with the KPZ fixed-point scaling function.}
    \label{fig:flow}
\end{figure}

\begin{figure}
    \begin{overpic}[width=0.45\textwidth]{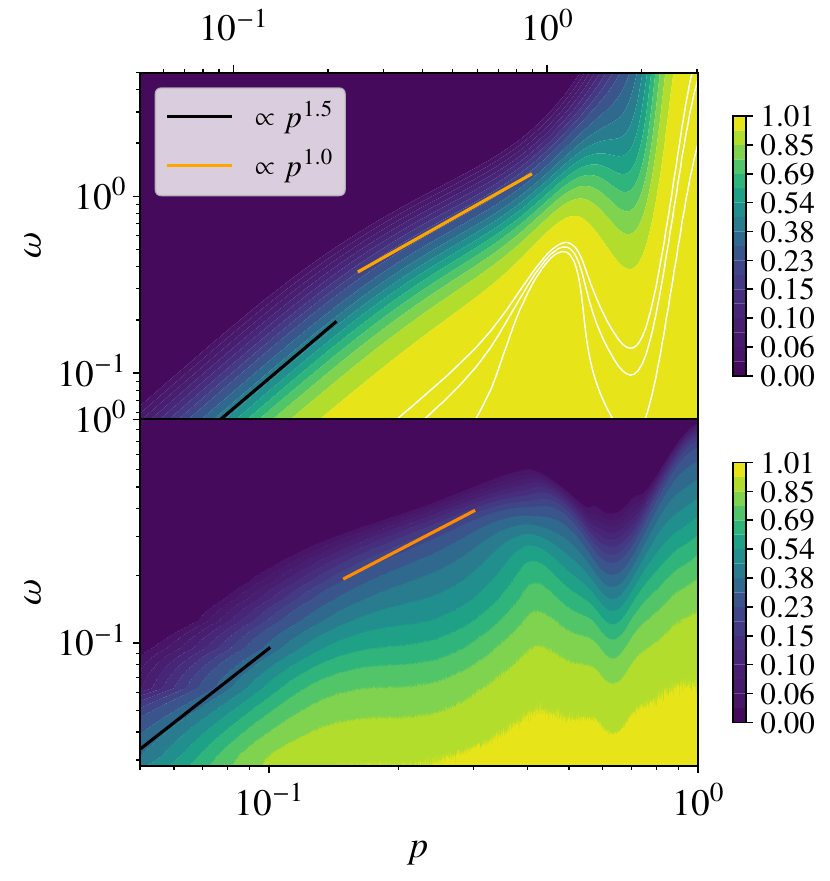}
        \put(0,215){(a)}
        \put(0,120){(b)}
    \end{overpic}
    \caption{Correlation function $C(\omega,p)/C(0,p)$ (a) from FRG; (b) from numerics. The KPZ and IB scaling regimes are indicated by the black and orange guidelines.}
    \label{fig:corromega}
\end{figure}

In practice, the flow equations are integrated numerically using a specific numerical scheme \cite{Benitez2009, Mathey2017, Gosteva2025KPZtwogrids} which allows one to describe not only the IR range, governed by the attractive KPZ fixed point, but  the whole range of wavenumbers and frequencies. This scheme, which involves two coupled grids, one  for dimensionless and one for dimensionful quantities, is described in details in Ref.~\cite{Gosteva2025KPZtwogrids}. It was shown for the KPZ equation to allow one to capture both UV (EW or IB) and IR (KPZ) scaling regimes.
 We use the regulators  $R^D_\kappa(p) = D_\kappa \hat r(\hat p^2)$ and
$R^\nu_\kappa(p) = A_\kappa \hat r(\hat p^2)$,
where
$\hat r(\hat p^2) = c \exp(-b\hat p^2) / \hat p^2$, with
$c=8$, $b=0.5$ \footnote{We have checked that the more common Wetterich regulator $\hat r(\hat p^2) = c / (\exp(\hat p^2)-1)$  yields the same behavior.}.
The typical evolution of the dimensionless functions $\hat f^D_\kappa$ and $\hat f^\nu_\kappa$ under the FRG flow is depicted in Fig.~\ref{fig:flow}.
 One observes that the effective viscosity, which can be read off from the value of $\hat{f}^\nu_\kappa$ at $\hat p=0$ as $\nu_\kappa=A_\kappa \hat{f}^\nu_\kappa(0,0)$ ($A_\kappa>0$),  starts from its negative initial value, and becomes positive during the flow. The two functions then reach a fixed identical form $\hat f^D_*=\hat f^\nu_*$ (displayed in the inset). This shows that the time-reversal symmetry, which is an exact property of the one-dimensional KPZ equation but is broken by the KS initial condition, emerges at large distances and times. Moreover, this fixed function is equal to the one obtained at the KPZ fixed point~\cite{Kloss2012}, which further identifies the IR state as the KPZ one.
The full correlation function (on the whole momentum range from UV to IR) is constructed from the dimensionful functions as $C(\omega,p) = 2  f^D_\kappa(\omega,p)/(\omega^2 + f^\nu_\kappa(\omega,p)^2 p^4)$ using the two-grid scheme. 
It is shown in Fig.~\ref{fig:corromega}(a) and commented along with the numerical results.

\begin{figure}
    \includegraphics[width=0.5\textwidth]{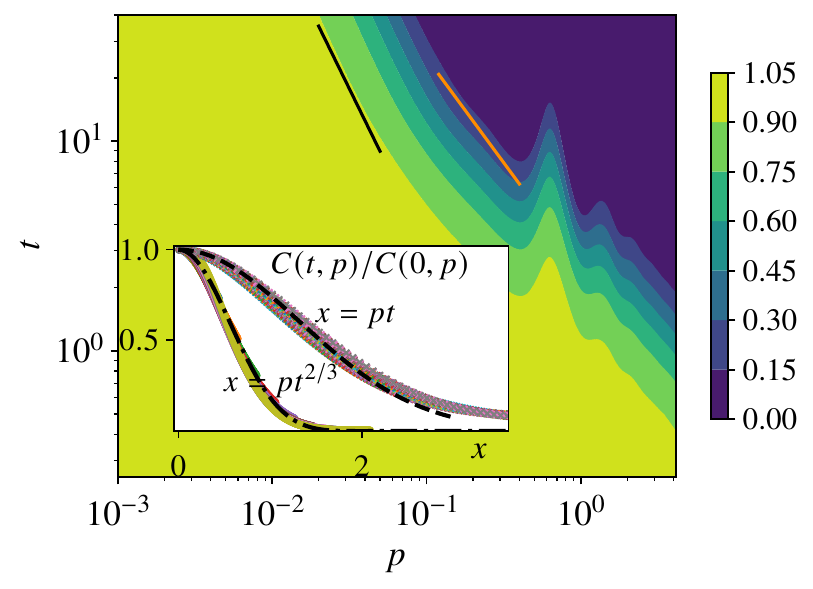}
    \caption{Two-point correlation function $C(t,p)/C(0,p)$ from DNS, plotted in logscales. The function $C(t,p)$ is computed for times $t=n/4$ with $n=0,1,2,\ldots$ from the simulations.
    The black line indicates the KPZ scaling $z=3/2$, and the orange one the IB scaling $z=1$. Inset: Extracted scaling functions, in the KPZ range ($p$ from 0.02 to 0.1, circles),  using the scaling variable $x=pt^{2/3}$; in the IB range ($p$ from 0.12 to 0.3, crosses),  using the scaling variable $x=pt$. They are compared with the exact Prah\"ofer-Spohn scaling function for the KPZ  fixed point (dash-dotted line), and with the exact asymptotic scaling function \eqref{eq:Casymp} from FRG for the IB fixed point (dashed line).
    }
    \label{fig:corrt}
\end{figure}

\noindent{\it Direct numerical simulations (DNS).} 
In our state-of-the-art simulations, we integrate the 1D KS equation \eqref{eq:KS} numerically using a pseudo-spectral method \cite{2007-canuto--zang} coupled with an exponential time differencing fourth order Runge-Kutta (ETDRK4) scheme for time-marching \cite{2002-cox-matthews, 2005-kassam-trefethen}. We overcome the challenges in numerical simulations for the 1D KS equation by implementing the ETDRK4 scheme with a CUDA C code which allows us to perform extensive simulations on computing clusters equipped with highly parallel NVIDIA Tesla P100 GPU processors. Our simulations involve a 1D periodic domain of size $L = 2^{17}$ on a grid of $2^{18}$ points. We generate $10^2$ independent realizations starting from random (Gaussian) initial data. For each of the realizations, we disregard the data for the initial time evolution and compute the spacetime correlation $C(t,p)$ by averaging over the steady state. Finally, averaging this correlation over different realizations yields the fully averaged data for $C(t,p)$ displayed in Fig.~\ref{fig:corrt}.

\noindent{\it Results.}
Let us first comment on the correlation function $C(t,p)$ computed from DNS.  At small momenta $p$, the contour lines exhibit a $z\simeq 3/2$ scaling, which is the one expected for KPZ universality. At larger momenta, over a range which extends up to the peak corresponding to the most unstable mode $k_0=\sqrt{|\nu|/(2\tau)}$, the slope of the contour lines decreases, evidencing another scaling regime, characterized by $z\lesssim 1$, which we identify with the IB scaling. The Fourier transform $C(\omega,p)$ is shown together with the one computed from FRG in Fig.~\ref{fig:corromega}. One observes that the two functions show a very similar structure.
In particular, both the KPZ scaling at small momenta and the IB scaling  in the intermediate range can be visualized in the contour lines. Note that in the FRG correlations, the largest-$p$ region with $z\sim 4$ reflects the initial condition and is non-universal.

\begin{figure}
    \begin{overpic}[percent,width=0.51\textwidth]{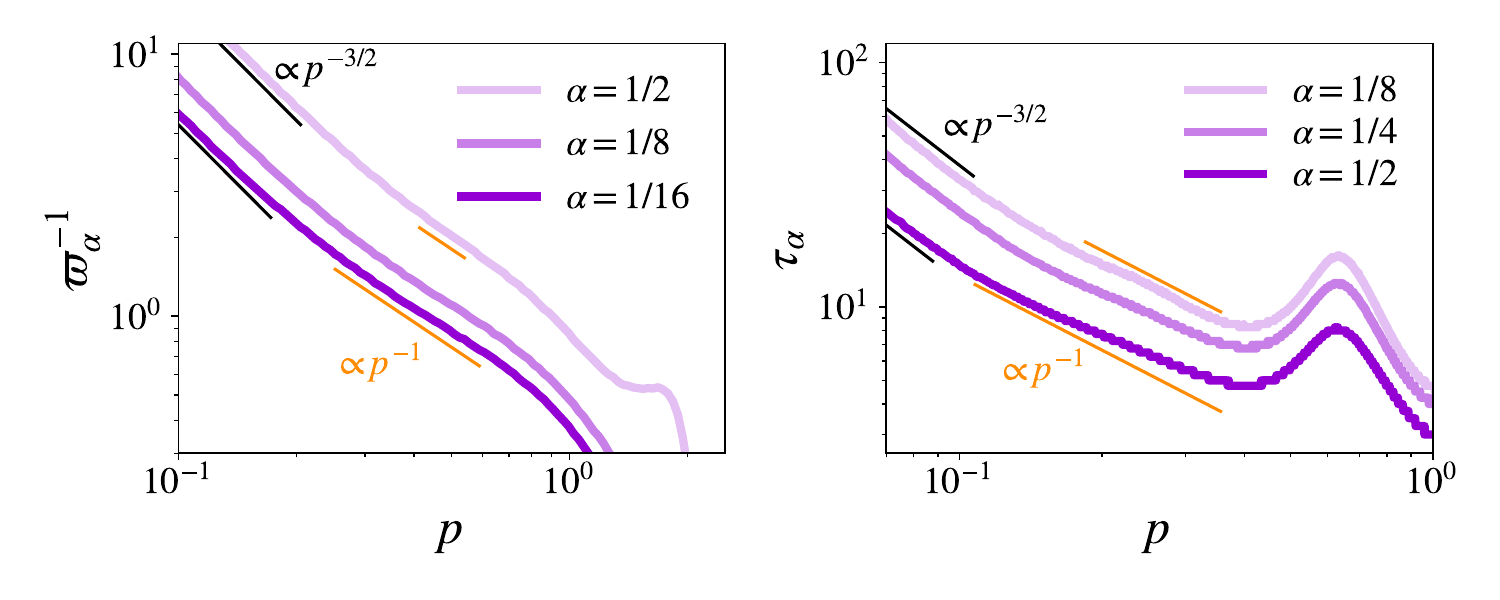}
    \put(2,35){\scriptsize(a)}
    \put(50,35){\scriptsize(b)}
    \end{overpic}
    \caption{(a) Inverse of the decorrelation frequencies $\varpi_{\alpha}(p)$ from FRG; (b)  decorrelation times $\tau_{\alpha}(p)$ from DNS (see text for definition). Both show the IB scaling $z=1$ over a range of intermediate momenta.}
    \label{fig:half-decay}
\end{figure}

We show in  Fig.~\ref{fig:half-decay} the decorrelation times $\tau_{\alpha}(p)$,
 defined as  $C(\tau_{\alpha}(p), p) = \alpha C(t=0,p)$ (DNS), or equivalently the decorrelation frequencies $\varpi_{\alpha}(p)$,
 defined as  $C(\varpi_{\alpha}(p), p) = \alpha C(\omega=0,p)$ (FRG), for different values of $\alpha$.
 One observes again that a region of momenta between  the IR modes and the peak exhibit the IB dynamical exponent $z\simeq 1$. In the DNS data, the exponent appears slightly smaller than one  for the larger $\alpha=1/2$, but still clearly distinct from $z=3/2$.  This $z\simeq 1$ scaling generically emerges, without any fine-tuning, as it originates from the vanishing of the effective viscosity. As a consequence, it is very robust. In particular, it appears irrespective of the IR regime, whether it is EW or KPZ, since it is observed already on small system sizes.

The scaling regimes can be further characterized by the whole scaling function, defined as $C(t,p)/C(0,p) = {\cal F}(pt^{1/z})$. For the KPZ fixed point, this scaling function was calculated exactly by Prah\"ofer and Spohn~\cite{Praehofer04}. For the IB fixed point,  an exact asymptotic form  was obtained within the FRG formalism \cite{Fontaine2023Unpredicted,Gosteva2025Burgers,Tarpin2018}. Its expression stems from (extended) symmetries and the limit of large wavenumbers, and is given by
\begin{equation}
 C(t,p)= C(0,p){\times}
    \begin{cases}
        \exp\left( - \mu_0 (pt)^2 \right),\, \text{small } t\,, \\
        \exp\left( - \mu_{\infty} p^2 |t| \right),\, \text{large } t
    \end{cases}
\label{eq:Casymp}
\end{equation} 
where $\mu_0$, $\mu_\infty$ are nonuniversal constants. 
The inset of Fig.~\ref{fig:corrt} shows the plot of $C(t,p)/C(0,p)$ obtained from DNS data selecting momenta $p$ lying in the KPZ region, and in the IB region. In both cases, the curves for different $p$ values collapse when plotted as a function of the corresponding scaling variable $pt^{1/z}$. Note that, although both are displayed on the same graph, the horizontal axis $x$ is different in each case, as $x=pt^{2/3}$ for momenta in the KPZ range, and $x=pt$ for momenta in the IB range. In the KPZ range, the scaling function obtained through the collapse precisely coincide with the exact one. In the IB range,
the obtained scaling function is well-fitted by a Gaussian curve, as predicted by the small-$t$ regime of \eqref{eq:Casymp}.  This is a further confirmation that the behavior of the correlation function over the whole intermediate range of momenta fully belongs to the IB universality class.
Note that at larger time delays, one observes hints of a slowing down of the decay of the scaling function, consistently with the crossover at large-$t$  predicted by \eqref{eq:Casymp}. However, the loss of accuracy of the data at larger time delays does not allow to fully resolve it. Let us emphasize that a similar behavior was predicted in the context of Navier-Stokes equation. In this case, the large-$t$ regime could be resolved and characterized in DNS data~\cite{Gorbunova2021scalar, Gosteva2025Euler}.

\noindent{\it Conclusions and perspectives.}
We have demonstrated, both from DNS and from FRG, the existence, in the correlations of the KS equation, of an extended and robust scaling regime characterized by the dynamical exponent $z=1$, which had been overlooked so far. This scaling regime controls the whole range of intermediate momenta, and it generically emerges  as a consequence of the KS dynamics. Indeed, it is rooted in the vanishing of the effective viscosity when changing from its microscopic KS negative value to its large-scale KPZ positive value. This behavior pertains to the IB universality class. In outlook, let us mention that in fact, beyond the IB regime, the approach to the  IR behavior, EW or KPZ, is more subtle than one might expect, and previous RG works deserve to be revisited with this respect. This will be investigated elsewhere.

\noindent{\it Acknowledgments.} NW thanks the LPMMC,  LG and LC the Instituto de F\'isica de la Facultad de Ingenier\'ia, for hospitality during the completion of this work, and they acknowledge support from the French-Uruguayan Institute of Physics (IFU$\Phi$). DR acknowledges support from Indian Institute of Science Education and Research Thiruvananthapuram (IISER TVM). LG acknowledges support  by the MSCA Cofund QuanG (Grant Number : 101081458) funded by the European Union. Views and opinions expressed are however those of the authors only and do not necessarily reflect those of the European Union or Université Grenoble Alpes. Neither the European Union nor the granting authority can be held responsible for them.


\begin{thebibliography}{47}%
\makeatletter
\providecommand \@ifxundefined [1]{%
 \@ifx{#1\undefined}
}%
\providecommand \@ifnum [1]{%
 \ifnum #1\expandafter \@firstoftwo
 \else \expandafter \@secondoftwo
 \fi
}%
\providecommand \@ifx [1]{%
 \ifx #1\expandafter \@firstoftwo
 \else \expandafter \@secondoftwo
 \fi
}%
\providecommand \natexlab [1]{#1}%
\providecommand \enquote  [1]{``#1''}%
\providecommand \bibnamefont  [1]{#1}%
\providecommand \bibfnamefont [1]{#1}%
\providecommand \citenamefont [1]{#1}%
\providecommand \href@noop [0]{\@secondoftwo}%
\providecommand \href [0]{\begingroup \@sanitize@url \@href}%
\providecommand \@href[1]{\@@startlink{#1}\@@href}%
\providecommand \@@href[1]{\endgroup#1\@@endlink}%
\providecommand \@sanitize@url [0]{\catcode `\\12\catcode `\$12\catcode
  `\&12\catcode `\#12\catcode `\^12\catcode `\_12\catcode `\%12\relax}%
\providecommand \@@startlink[1]{}%
\providecommand \@@endlink[0]{}%
\providecommand \url  [0]{\begingroup\@sanitize@url \@url }%
\providecommand \@url [1]{\endgroup\@href {#1}{\urlprefix }}%
\providecommand \urlprefix  [0]{URL }%
\providecommand \Eprint [0]{\href }%
\providecommand \doibase [0]{https://doi.org/}%
\providecommand \selectlanguage [0]{\@gobble}%
\providecommand \bibinfo  [0]{\@secondoftwo}%
\providecommand \bibfield  [0]{\@secondoftwo}%
\providecommand \translation [1]{[#1]}%
\providecommand \BibitemOpen [0]{}%
\providecommand \bibitemStop [0]{}%
\providecommand \bibitemNoStop [0]{.\EOS\space}%
\providecommand \EOS [0]{\spacefactor3000\relax}%
\providecommand \BibitemShut  [1]{\csname bibitem#1\endcsname}%
\let\auto@bib@innerbib\@empty
\bibitem [{\citenamefont {Kuramoto}\ and\ \citenamefont
  {Tsuzuki}(1975)}]{KuramotoTsuzuki1975}%
  \BibitemOpen
  \bibfield  {author} {\bibinfo {author} {\bibfnamefont {Y.}~\bibnamefont
  {Kuramoto}}\ and\ \bibinfo {author} {\bibfnamefont {T.}~\bibnamefont
  {Tsuzuki}},\ }\bibfield  {title} {\bibinfo {title} {On the formation of
  dissipative structures in reaction-diffusion systems},\ }\href
  {https://doi.org/https://doi.org/10.1143/PTP.54.687} {\bibfield  {journal}
  {\bibinfo  {journal} {Progress of Theoretical Physics}\ }\textbf {\bibinfo
  {volume} {54}},\ \bibinfo {pages} {687} (\bibinfo {year} {1975})}\BibitemShut
  {NoStop}%
\bibitem [{\citenamefont {Kuramoto}\ and\ \citenamefont
  {Tsuzuki}(1976)}]{KuramotoTsuzuki1976}%
  \BibitemOpen
  \bibfield  {author} {\bibinfo {author} {\bibfnamefont {Y.}~\bibnamefont
  {Kuramoto}}\ and\ \bibinfo {author} {\bibfnamefont {T.}~\bibnamefont
  {Tsuzuki}},\ }\bibfield  {title} {\bibinfo {title} {Persistent propagation of
  concentration waves in dissipative media far from thermal equilibrium},\
  }\href@noop {} {\bibfield  {journal} {\bibinfo  {journal} {Progress of
  Theoretical Physics}\ }\textbf {\bibinfo {volume} {55}},\ \bibinfo {pages}
  {356} (\bibinfo {year} {1976})}\BibitemShut {NoStop}%
\bibitem [{\citenamefont {Kuramoto}(1978)}]{Kuramoto1978}%
  \BibitemOpen
  \bibfield  {author} {\bibinfo {author} {\bibfnamefont {Y.}~\bibnamefont
  {Kuramoto}},\ }\bibfield  {title} {\bibinfo {title} {Diffusion-induced chaos
  in reaction systems},\ }\href@noop {} {\bibfield  {journal} {\bibinfo
  {journal} {Supplement of Progress of Theoretical Physics}\ }\textbf {\bibinfo
  {volume} {64}},\ \bibinfo {pages} {346} (\bibinfo {year} {1978})}\BibitemShut
  {NoStop}%
\bibitem [{\citenamefont
  {Sivashinsky}(1977{\natexlab{a}})}]{Sivashinsky1977_1}%
  \BibitemOpen
  \bibfield  {author} {\bibinfo {author} {\bibfnamefont {G.}~\bibnamefont
  {Sivashinsky}},\ }\bibfield  {title} {\bibinfo {title} {Nonlinear analysis of
  hydrodynamic instability in laminar flames. {Part I}. {D}erivation of basic
  equations},\ }\href@noop {} {\bibfield  {journal} {\bibinfo  {journal} {Acta
  Astronautica}\ }\textbf {\bibinfo {volume} {4}},\ \bibinfo {pages} {1177}
  (\bibinfo {year} {1977}{\natexlab{a}})}\BibitemShut {NoStop}%
\bibitem [{\citenamefont {Michelson}\ and\ \citenamefont
  {Sivashinsky}(1977)}]{MichelsonSivashinsky1977_2}%
  \BibitemOpen
  \bibfield  {author} {\bibinfo {author} {\bibfnamefont {D.~M.}\ \bibnamefont
  {Michelson}}\ and\ \bibinfo {author} {\bibfnamefont {G.~I.}\ \bibnamefont
  {Sivashinsky}},\ }\bibfield  {title} {\bibinfo {title} {Nonlinear analysis of
  hydrodynamic instability in laminar flames. {Part II}. {N}umerical
  experiments},\ }\href@noop {} {\bibfield  {journal} {\bibinfo  {journal}
  {Acta Astronautica}\ }\textbf {\bibinfo {volume} {4}},\ \bibinfo {pages}
  {1207} (\bibinfo {year} {1977})}\BibitemShut {NoStop}%
\bibitem [{\citenamefont {Sivashinsky}(1977{\natexlab{b}})}]{Sivashinsky1977}%
  \BibitemOpen
  \bibfield  {author} {\bibinfo {author} {\bibfnamefont {G.}~\bibnamefont
  {Sivashinsky}},\ }\bibfield  {title} {\bibinfo {title} {On self-turbulization
  of a laminar flame},\ }\href@noop {} {\bibfield  {journal} {\bibinfo
  {journal} {Acta Astronautica}\ }\textbf {\bibinfo {volume} {6}},\ \bibinfo
  {pages} {569} (\bibinfo {year} {1977}{\natexlab{b}})}\BibitemShut {NoStop}%
\bibitem [{\citenamefont {Sivashinsky}(1980)}]{Sivashinsky1980}%
  \BibitemOpen
  \bibfield  {author} {\bibinfo {author} {\bibfnamefont {G.}~\bibnamefont
  {Sivashinsky}},\ }\bibfield  {title} {\bibinfo {title} {On flame propagation
  under conditions of stoichiometry},\ }\href@noop {} {\bibfield  {journal}
  {\bibinfo  {journal} {SIAM Journal on Applied Mathematics}\ }\textbf
  {\bibinfo {volume} {39}},\ \bibinfo {pages} {67} (\bibinfo {year}
  {1980})}\BibitemShut {NoStop}%
\bibitem [{\citenamefont {Nepomnyashchii}(1974)}]{Nepomnyashchii1974}%
  \BibitemOpen
  \bibfield  {author} {\bibinfo {author} {\bibfnamefont {A.}~\bibnamefont
  {Nepomnyashchii}},\ }\bibfield  {title} {\bibinfo {title} {Stability of wavy
  conditions in a film flowing down an inclined plan},\ }\href
  {https://doi.org/https://doi.org/10.1007/BF01025515} {\bibfield  {journal}
  {\bibinfo  {journal} {Fluid Dynamics}\ }\textbf {\bibinfo {volume} {9}},\
  \bibinfo {pages} {354–359} (\bibinfo {year} {1974})}\BibitemShut {NoStop}%
\bibitem [{\citenamefont {Sivashinsky}\ and\ \citenamefont
  {Michelson}(1980)}]{SivashinskyMichelson1980}%
  \BibitemOpen
  \bibfield  {author} {\bibinfo {author} {\bibfnamefont {G.~I.}\ \bibnamefont
  {Sivashinsky}}\ and\ \bibinfo {author} {\bibfnamefont {D.~M.}\ \bibnamefont
  {Michelson}},\ }\bibfield  {title} {\bibinfo {title} {On irregular wavy flow
  of a liquid film down a vertical plane},\ }\href
  {https://doi.org/10.1143/PTP.63.2112} {\bibfield  {journal} {\bibinfo
  {journal} {Progress of Theoretical Physics}\ }\textbf {\bibinfo {volume}
  {63}},\ \bibinfo {pages} {2112} (\bibinfo {year} {1980})}\BibitemShut
  {NoStop}%
\bibitem [{\citenamefont {Chen}\ and\ \citenamefont {Chang}(1986)}]{Chen1986}%
  \BibitemOpen
  \bibfield  {author} {\bibinfo {author} {\bibfnamefont {L.-H.}\ \bibnamefont
  {Chen}}\ and\ \bibinfo {author} {\bibfnamefont {H.-C.}\ \bibnamefont
  {Chang}},\ }\bibfield  {title} {\bibinfo {title} {Nonlinear waves on liquid
  film surfaces—{II}. {B}ifurcation analyses of the long-wave equation},\
  }\href {https://doi.org/https://doi.org/10.1016/0009-2509(86)80033-1}
  {\bibfield  {journal} {\bibinfo  {journal} {Chemical Engineering Science}\
  }\textbf {\bibinfo {volume} {41}},\ \bibinfo {pages} {2477} (\bibinfo {year}
  {1986})}\BibitemShut {NoStop}%
\bibitem [{\citenamefont {La~Quey}\ \emph {et~al.}(1975)\citenamefont
  {La~Quey}, \citenamefont {Mahajan}, \citenamefont {Rutherford},\ and\
  \citenamefont {Tang}}]{LaQuey1975}%
  \BibitemOpen
  \bibfield  {author} {\bibinfo {author} {\bibfnamefont {R.~E.}\ \bibnamefont
  {La~Quey}}, \bibinfo {author} {\bibfnamefont {S.~M.}\ \bibnamefont
  {Mahajan}}, \bibinfo {author} {\bibfnamefont {P.~H.}\ \bibnamefont
  {Rutherford}},\ and\ \bibinfo {author} {\bibfnamefont {W.~M.}\ \bibnamefont
  {Tang}},\ }\bibfield  {title} {\bibinfo {title} {Nonlinear saturation of the
  trapped-ion mode},\ }\href@noop {} {\bibfield  {journal} {\bibinfo  {journal}
  {Physical Review Letters}\ }\textbf {\bibinfo {volume} {34}},\ \bibinfo
  {pages} {391} (\bibinfo {year} {1975})}\BibitemShut {NoStop}%
\bibitem [{\citenamefont {Cuerno}\ \emph {et~al.}(1995)\citenamefont {Cuerno},
  \citenamefont {Makse}, \citenamefont {Tomassone}, \citenamefont
  {Harrington},\ and\ \citenamefont {Stanley}}]{Cuerno1995}%
  \BibitemOpen
  \bibfield  {author} {\bibinfo {author} {\bibfnamefont {R.}~\bibnamefont
  {Cuerno}}, \bibinfo {author} {\bibfnamefont {H.~A.}\ \bibnamefont {Makse}},
  \bibinfo {author} {\bibfnamefont {S.}~\bibnamefont {Tomassone}}, \bibinfo
  {author} {\bibfnamefont {S.}~\bibnamefont {Harrington}},\ and\ \bibinfo
  {author} {\bibfnamefont {H.~E.}\ \bibnamefont {Stanley}},\ }\bibfield
  {title} {\bibinfo {title} {Dynamic scaling of ion-sputtered surfaces},\
  }\href@noop {} {\bibfield  {journal} {\bibinfo  {journal} {Physical Review
  Letters}\ }\textbf {\bibinfo {volume} {75}},\ \bibinfo {pages} {4464}
  (\bibinfo {year} {1995})}\BibitemShut {NoStop}%
\bibitem [{\citenamefont {Kanjilal}\ and\ \citenamefont
  {Kanjilal}(2006)}]{Kanjilal2006}%
  \BibitemOpen
  \bibfield  {author} {\bibinfo {author} {\bibfnamefont {A.}~\bibnamefont
  {Kanjilal}}\ and\ \bibinfo {author} {\bibfnamefont {D.}~\bibnamefont
  {Kanjilal}},\ }\bibfield  {title} {\bibinfo {title} {Surface roughening in
  {Si}$_{1-x}${Ge}$_x$ alloy films by 100 {MeV} {Au}: {C}omposition
  dependency},\ }\href@noop {} {\bibfield  {journal} {\bibinfo  {journal}
  {Solid State Communications}\ }\textbf {\bibinfo {volume} {139}},\ \bibinfo
  {pages} {531} (\bibinfo {year} {2006})}\BibitemShut {NoStop}%
\bibitem [{\citenamefont {Losert}\ \emph {et~al.}(1998)\citenamefont {Losert},
  \citenamefont {Shi}, \citenamefont {Cummins},\ and\ \citenamefont
  {Cannell}}]{Losert1998}%
  \BibitemOpen
  \bibfield  {author} {\bibinfo {author} {\bibfnamefont {W.}~\bibnamefont
  {Losert}}, \bibinfo {author} {\bibfnamefont {B.~Q.}\ \bibnamefont {Shi}},
  \bibinfo {author} {\bibfnamefont {H.~Z.}\ \bibnamefont {Cummins}},\ and\
  \bibinfo {author} {\bibfnamefont {D.~S.}\ \bibnamefont {Cannell}},\
  }\bibfield  {title} {\bibinfo {title} {Spatiotemporal chaos in surface
  waves},\ }\href@noop {} {\bibfield  {journal} {\bibinfo  {journal}
  {Proceedings of the National Academy of Sciences of the USA}\ }\textbf
  {\bibinfo {volume} {95}},\ \bibinfo {pages} {431} (\bibinfo {year}
  {1998})}\BibitemShut {NoStop}%
\bibitem [{\citenamefont {Grinstein}\ \emph {et~al.}(1996)\citenamefont
  {Grinstein}, \citenamefont {Jayaprakash},\ and\ \citenamefont
  {Pandit}}]{Grinstein1996}%
  \BibitemOpen
  \bibfield  {author} {\bibinfo {author} {\bibfnamefont {G.}~\bibnamefont
  {Grinstein}}, \bibinfo {author} {\bibfnamefont {C.}~\bibnamefont
  {Jayaprakash}},\ and\ \bibinfo {author} {\bibfnamefont {R.}~\bibnamefont
  {Pandit}},\ }\bibfield  {title} {\bibinfo {title} {Conjectures about phase
  turbulence in the complex ginzburg-landau equation},\ }\href
  {https://doi.org/https://doi.org/10.1016/0167-2789(95)00036-4} {\bibfield
  {journal} {\bibinfo  {journal} {Physica D: Nonlinear Phenomena}\ }\textbf
  {\bibinfo {volume} {90}},\ \bibinfo {pages} {96} (\bibinfo {year}
  {1996})}\BibitemShut {NoStop}%
\bibitem [{\citenamefont {Aranson}\ and\ \citenamefont
  {Kramer}(2002)}]{Aranson2002}%
  \BibitemOpen
  \bibfield  {author} {\bibinfo {author} {\bibfnamefont {I.~S.}\ \bibnamefont
  {Aranson}}\ and\ \bibinfo {author} {\bibfnamefont {L.}~\bibnamefont
  {Kramer}},\ }\bibfield  {title} {\bibinfo {title} {The world of the complex
  {G}inzburg-{L}andau equation},\ }\href
  {https://doi.org/10.1103/revmodphys.74.99} {\bibfield  {journal} {\bibinfo
  {journal} {Reviews of Modern Physics}\ }\textbf {\bibinfo {volume} {74}},\
  \bibinfo {pages} {99} (\bibinfo {year} {2002})}\BibitemShut {NoStop}%
\bibitem [{\citenamefont {Kardar}\ \emph {et~al.}(1986)\citenamefont {Kardar},
  \citenamefont {Parisi},\ and\ \citenamefont {Zhang}}]{Kardar1986}%
  \BibitemOpen
  \bibfield  {author} {\bibinfo {author} {\bibfnamefont {M.}~\bibnamefont
  {Kardar}}, \bibinfo {author} {\bibfnamefont {G.}~\bibnamefont {Parisi}},\
  and\ \bibinfo {author} {\bibfnamefont {Y.-C.}\ \bibnamefont {Zhang}},\
  }\bibfield  {title} {\bibinfo {title} {Dynamic scaling of growing
  interfaces},\ }\href {https://doi.org/10.1103/PhysRevLett.56.889} {\bibfield
  {journal} {\bibinfo  {journal} {Phys. Rev. Lett.}\ }\textbf {\bibinfo
  {volume} {56}},\ \bibinfo {pages} {889} (\bibinfo {year} {1986})}\BibitemShut
  {NoStop}%
\bibitem [{\citenamefont {Fujisaka}\ and\ \citenamefont
  {Yamada}(1977)}]{Fujisaka1977}%
  \BibitemOpen
  \bibfield  {author} {\bibinfo {author} {\bibfnamefont {H.}~\bibnamefont
  {Fujisaka}}\ and\ \bibinfo {author} {\bibfnamefont {T.}~\bibnamefont
  {Yamada}},\ }\bibfield  {title} {\bibinfo {title} {Theoretical study of a
  chemical turbulence},\ }\href {https://doi.org/10.1143/PTP.57.734} {\bibfield
   {journal} {\bibinfo  {journal} {Progress of Theoretical Physics}\ }\textbf
  {\bibinfo {volume} {57}},\ \bibinfo {pages} {734} (\bibinfo {year}
  {1977})}\BibitemShut {NoStop}%
\bibitem [{\citenamefont {Yakhot}(1981)}]{Yakhot1980}%
  \BibitemOpen
  \bibfield  {author} {\bibinfo {author} {\bibfnamefont {V.}~\bibnamefont
  {Yakhot}},\ }\bibfield  {title} {\bibinfo {title} {Large-scale properties of
  unstable systems governed by the {K}uramoto-{S}ivashinsky equation},\ }\href
  {https://doi.org/10.1103/PhysRevA.24.642} {\bibfield  {journal} {\bibinfo
  {journal} {Phys. Rev. A}\ }\textbf {\bibinfo {volume} {24}},\ \bibinfo
  {pages} {642} (\bibinfo {year} {1981})}\BibitemShut {NoStop}%
\bibitem [{\citenamefont {Ueno}\ \emph {et~al.}(2005)\citenamefont {Ueno},
  \citenamefont {Sakaguchi},\ and\ \citenamefont {Okamura}}]{Ueno2005}%
  \BibitemOpen
  \bibfield  {author} {\bibinfo {author} {\bibfnamefont {K.}~\bibnamefont
  {Ueno}}, \bibinfo {author} {\bibfnamefont {H.}~\bibnamefont {Sakaguchi}},\
  and\ \bibinfo {author} {\bibfnamefont {M.}~\bibnamefont {Okamura}},\
  }\bibfield  {title} {\bibinfo {title} {Renormalization-group and numerical
  analysis of a noisy {K}uramoto-{S}ivashinsky equation in 1+1 dimensions},\
  }\href {https://doi.org/10.1103/PhysRevE.71.046138} {\bibfield  {journal}
  {\bibinfo  {journal} {Phys. Rev. E}\ }\textbf {\bibinfo {volume} {71}},\
  \bibinfo {pages} {046138} (\bibinfo {year} {2005})}\BibitemShut {NoStop}%
\bibitem [{\citenamefont {Zaleski}(1989)}]{Zaleski1989}%
  \BibitemOpen
  \bibfield  {author} {\bibinfo {author} {\bibfnamefont {S.}~\bibnamefont
  {Zaleski}},\ }\bibfield  {title} {\bibinfo {title} {A stochastic model for
  the large scale dynamics of some fluctuating interfaces},\ }\href
  {https://doi.org/https://doi.org/10.1016/0167-2789(89)90266-2} {\bibfield
  {journal} {\bibinfo  {journal} {Physica D: Nonlinear Phenomena}\ }\textbf
  {\bibinfo {volume} {34}},\ \bibinfo {pages} {427} (\bibinfo {year}
  {1989})}\BibitemShut {NoStop}%
\bibitem [{\citenamefont {Roy}\ and\ \citenamefont {Pandit}(2020)}]{Roy2020}%
  \BibitemOpen
  \bibfield  {author} {\bibinfo {author} {\bibfnamefont {D.}~\bibnamefont
  {Roy}}\ and\ \bibinfo {author} {\bibfnamefont {R.}~\bibnamefont {Pandit}},\
  }\bibfield  {title} {\bibinfo {title} {One-dimensional {Kardar-Parisi-Zhang}
  and {Kuramoto-Sivashinsky} universality class: {L}imit distributions},\
  }\href {https://doi.org/10.1103/PhysRevE.101.030103} {\bibfield  {journal}
  {\bibinfo  {journal} {Phys. Rev. E}\ }\textbf {\bibinfo {volume} {101}},\
  \bibinfo {pages} {030103(R)} (\bibinfo {year} {2020})}\BibitemShut {NoStop}%
\bibitem [{\citenamefont {Hyman}\ \emph {et~al.}(1986)\citenamefont {Hyman},
  \citenamefont {Nicolaenko},\ and\ \citenamefont {Zaleski}}]{Hyman1986}%
  \BibitemOpen
  \bibfield  {author} {\bibinfo {author} {\bibfnamefont {J.~M.}\ \bibnamefont
  {Hyman}}, \bibinfo {author} {\bibfnamefont {B.}~\bibnamefont {Nicolaenko}},\
  and\ \bibinfo {author} {\bibfnamefont {S.}~\bibnamefont {Zaleski}},\
  }\bibfield  {title} {\bibinfo {title} {Order and complexity in the
  {Kuramoto-Sivashinsky} model of weakly turbulent interfaces},\ }\href
  {https://doi.org/https://doi.org/10.1016/0167-2789(86)90136-3} {\bibfield
  {journal} {\bibinfo  {journal} {Physica D: Nonlinear Phenomena}\ }\textbf
  {\bibinfo {volume} {23}},\ \bibinfo {pages} {265} (\bibinfo {year}
  {1986})}\BibitemShut {NoStop}%
\bibitem [{\citenamefont {Sneppen}\ \emph {et~al.}(1992)\citenamefont
  {Sneppen}, \citenamefont {Krug}, \citenamefont {Jensen}, \citenamefont
  {Jayaprakash},\ and\ \citenamefont {Bohr}}]{Sneppen1992}%
  \BibitemOpen
  \bibfield  {author} {\bibinfo {author} {\bibfnamefont {K.}~\bibnamefont
  {Sneppen}}, \bibinfo {author} {\bibfnamefont {J.}~\bibnamefont {Krug}},
  \bibinfo {author} {\bibfnamefont {M.~H.}\ \bibnamefont {Jensen}}, \bibinfo
  {author} {\bibfnamefont {C.}~\bibnamefont {Jayaprakash}},\ and\ \bibinfo
  {author} {\bibfnamefont {T.}~\bibnamefont {Bohr}},\ }\bibfield  {title}
  {\bibinfo {title} {Dynamic scaling and crossover analysis for the
  {K}uramoto-{S}ivashinsky equation},\ }\href
  {https://doi.org/10.1103/PhysRevA.46.R7351} {\bibfield  {journal} {\bibinfo
  {journal} {Phys. Rev. A}\ }\textbf {\bibinfo {volume} {46}},\ \bibinfo
  {pages} {R7351} (\bibinfo {year} {1992})}\BibitemShut {NoStop}%
\bibitem [{\citenamefont {Hayot}\ \emph {et~al.}(1993)\citenamefont {Hayot},
  \citenamefont {Jayaprakash},\ and\ \citenamefont {Josserand}}]{Hayot1993}%
  \BibitemOpen
  \bibfield  {author} {\bibinfo {author} {\bibfnamefont {F.}~\bibnamefont
  {Hayot}}, \bibinfo {author} {\bibfnamefont {C.}~\bibnamefont {Jayaprakash}},\
  and\ \bibinfo {author} {\bibfnamefont {C.}~\bibnamefont {Josserand}},\
  }\bibfield  {title} {\bibinfo {title} {Long-wavelength properties of the
  {Kuramoto-Sivashinsky} equation},\ }\href
  {https://doi.org/10.1103/PhysRevE.47.911} {\bibfield  {journal} {\bibinfo
  {journal} {Phys. Rev. E}\ }\textbf {\bibinfo {volume} {47}},\ \bibinfo
  {pages} {911} (\bibinfo {year} {1993})}\BibitemShut {NoStop}%
\bibitem [{\citenamefont {Fontaine}\ \emph {et~al.}(2023)\citenamefont
  {Fontaine}, \citenamefont {Vercesi}, \citenamefont {Brachet},\ and\
  \citenamefont {Canet}}]{Fontaine2023Unpredicted}%
  \BibitemOpen
  \bibfield  {author} {\bibinfo {author} {\bibfnamefont {C.}~\bibnamefont
  {Fontaine}}, \bibinfo {author} {\bibfnamefont {F.}~\bibnamefont {Vercesi}},
  \bibinfo {author} {\bibfnamefont {M.}~\bibnamefont {Brachet}},\ and\ \bibinfo
  {author} {\bibfnamefont {L.}~\bibnamefont {Canet}},\ }\bibfield  {title}
  {\bibinfo {title} {Unpredicted scaling of the one-dimensional
  {K}ardar-{P}arisi-{Z}hang equation},\ }\href
  {https://doi.org/10.1103/PhysRevLett.131.247101} {\bibfield  {journal}
  {\bibinfo  {journal} {Phys. Rev. Lett.}\ }\textbf {\bibinfo {volume} {131}},\
  \bibinfo {pages} {247101} (\bibinfo {year} {2023})}\BibitemShut {NoStop}%
\bibitem [{\citenamefont {Gosteva}\ \emph {et~al.}(2024)\citenamefont
  {Gosteva}, \citenamefont {Tarpin}, \citenamefont {Wschebor},\ and\
  \citenamefont {Canet}}]{Gosteva2025Burgers}%
  \BibitemOpen
  \bibfield  {author} {\bibinfo {author} {\bibfnamefont {L.}~\bibnamefont
  {Gosteva}}, \bibinfo {author} {\bibfnamefont {M.}~\bibnamefont {Tarpin}},
  \bibinfo {author} {\bibfnamefont {N.}~\bibnamefont {Wschebor}},\ and\
  \bibinfo {author} {\bibfnamefont {L.}~\bibnamefont {Canet}},\ }\bibfield
  {title} {\bibinfo {title} {Inviscid fixed point of the multidimensional
  {B}urgers--{K}ardar-{P}arisi-{Z}hang equation},\ }\href
  {https://doi.org/10.1103/PhysRevE.110.054118} {\bibfield  {journal} {\bibinfo
   {journal} {Phys. Rev. E}\ }\textbf {\bibinfo {volume} {110}},\ \bibinfo
  {pages} {054118} (\bibinfo {year} {2024})}\BibitemShut {NoStop}%
\bibitem [{\citenamefont {Gosteva}\ \emph
  {et~al.}(2025{\natexlab{a}})\citenamefont {Gosteva}, \citenamefont
  {Wschebor},\ and\ \citenamefont {Canet}}]{Gosteva2025KPZtwogrids}%
  \BibitemOpen
  \bibfield  {author} {\bibinfo {author} {\bibfnamefont {L.}~\bibnamefont
  {Gosteva}}, \bibinfo {author} {\bibfnamefont {N.}~\bibnamefont {Wschebor}},\
  and\ \bibinfo {author} {\bibfnamefont {L.}~\bibnamefont {Canet}},\ }\bibfield
   {title} {\bibinfo {title} {Unveiling the different scaling regimes of the
  one-dimensional {Kardar–Parisi–Zhang–Burgers} equation using the
  functional renormalisation group},\ }\href
  {https://doi.org/10.1088/1742-5468/ae1b85} {\bibfield  {journal} {\bibinfo
  {journal} {Journal of Statistical Mechanics: Theory and Experiment}\ }\textbf
  {\bibinfo {volume} {2025}},\ \bibinfo {pages} {114002} (\bibinfo {year}
  {2025}{\natexlab{a}})}\BibitemShut {NoStop}%
\bibitem [{\citenamefont {Vercesi}\ \emph {et~al.}(2024)\citenamefont
  {Vercesi}, \citenamefont {Poirier}, \citenamefont {Minguzzi},\ and\
  \citenamefont {Canet}}]{Vercesi2024}%
  \BibitemOpen
  \bibfield  {author} {\bibinfo {author} {\bibfnamefont {F.}~\bibnamefont
  {Vercesi}}, \bibinfo {author} {\bibfnamefont {S.}~\bibnamefont {Poirier}},
  \bibinfo {author} {\bibfnamefont {A.}~\bibnamefont {Minguzzi}},\ and\
  \bibinfo {author} {\bibfnamefont {L.}~\bibnamefont {Canet}},\ }\bibfield
  {title} {\bibinfo {title} {Scaling regimes of the one-dimensional phase
  turbulence in the deterministic complex {G}inzburg-{L}andau equation},\
  }\href {https://doi.org/10.1103/PhysRevE.109.064149} {\bibfield  {journal}
  {\bibinfo  {journal} {Phys. Rev. E}\ }\textbf {\bibinfo {volume} {109}},\
  \bibinfo {pages} {064149} (\bibinfo {year} {2024})}\BibitemShut {NoStop}%
\bibitem [{\citenamefont {Wetterich}(1993)}]{Wetterich1993}%
  \BibitemOpen
  \bibfield  {author} {\bibinfo {author} {\bibfnamefont {C.}~\bibnamefont
  {Wetterich}},\ }\bibfield  {title} {\bibinfo {title} {Exact evolution
  equation for the effective potential},\ }\href
  {https://doi.org/https://doi.org/10.1016/0370-2693(93)90726-X} {\bibfield
  {journal} {\bibinfo  {journal} {Physics Letters B}\ }\textbf {\bibinfo
  {volume} {301}},\ \bibinfo {pages} {90} (\bibinfo {year} {1993})}\BibitemShut
  {NoStop}%
\bibitem [{\citenamefont {Delamotte}(2012)}]{Delamotte2012}%
  \BibitemOpen
  \bibfield  {author} {\bibinfo {author} {\bibfnamefont {B.}~\bibnamefont
  {Delamotte}},\ }\bibinfo {title} {An introduction to the nonperturbative
  renormalization group},\ in\ \href
  {https://doi.org/10.1007/978-3-642-27320-9_2} {\emph {\bibinfo {booktitle}
  {Renormalization Group and Effective Field Theory Approaches to Many-Body
  Systems}}},\ \bibinfo {editor} {edited by\ \bibinfo {editor} {\bibfnamefont
  {A.}~\bibnamefont {Schwenk}}\ and\ \bibinfo {editor} {\bibfnamefont
  {J.}~\bibnamefont {Polonyi}}}\ (\bibinfo  {publisher} {Springer Berlin
  Heidelberg},\ \bibinfo {address} {Berlin, Heidelberg},\ \bibinfo {year}
  {2012})\ pp.\ \bibinfo {pages} {49--132}\BibitemShut {NoStop}%
\bibitem [{\citenamefont {Dupuis}\ \emph {et~al.}(2021)\citenamefont {Dupuis},
  \citenamefont {Canet}, \citenamefont {Eichhorn}, \citenamefont {Metzner},
  \citenamefont {Pawlowski}, \citenamefont {Tissier},\ and\ \citenamefont
  {Wschebor}}]{Dupuis2021}%
  \BibitemOpen
  \bibfield  {author} {\bibinfo {author} {\bibfnamefont {N.}~\bibnamefont
  {Dupuis}}, \bibinfo {author} {\bibfnamefont {L.}~\bibnamefont {Canet}},
  \bibinfo {author} {\bibfnamefont {A.}~\bibnamefont {Eichhorn}}, \bibinfo
  {author} {\bibfnamefont {W.}~\bibnamefont {Metzner}}, \bibinfo {author}
  {\bibfnamefont {J.}~\bibnamefont {Pawlowski}}, \bibinfo {author}
  {\bibfnamefont {M.}~\bibnamefont {Tissier}},\ and\ \bibinfo {author}
  {\bibfnamefont {N.}~\bibnamefont {Wschebor}},\ }\bibfield  {title} {\bibinfo
  {title} {The nonperturbative functional renormalization group and its
  applications},\ }\href
  {https://doi.org/https://doi.org/10.1016/j.physrep.2021.01.001} {\bibfield
  {journal} {\bibinfo  {journal} {Physics Reports}\ }\textbf {\bibinfo {volume}
  {910}},\ \bibinfo {pages} {1} (\bibinfo {year} {2021})}\BibitemShut {NoStop}%
\bibitem [{\citenamefont {Martin}\ \emph {et~al.}(1973)\citenamefont {Martin},
  \citenamefont {Siggia},\ and\ \citenamefont {Rose}}]{Martin1973}%
  \BibitemOpen
  \bibfield  {author} {\bibinfo {author} {\bibfnamefont {P.~C.}\ \bibnamefont
  {Martin}}, \bibinfo {author} {\bibfnamefont {E.~D.}\ \bibnamefont {Siggia}},\
  and\ \bibinfo {author} {\bibfnamefont {H.~A.}\ \bibnamefont {Rose}},\
  }\bibfield  {title} {\bibinfo {title} {Statistical dynamics of classical
  systems},\ }\href {https://doi.org/10.1103/PhysRevA.8.423} {\bibfield
  {journal} {\bibinfo  {journal} {Phys. Rev. A}\ }\textbf {\bibinfo {volume}
  {8}},\ \bibinfo {pages} {423} (\bibinfo {year} {1973})}\BibitemShut {NoStop}%
\bibitem [{\citenamefont {Janssen}(1976)}]{Janssen1976}%
  \BibitemOpen
  \bibfield  {author} {\bibinfo {author} {\bibfnamefont {H.-K.}\ \bibnamefont
  {Janssen}},\ }\bibfield  {title} {\bibinfo {title} {On a lagrangean for
  classical field dynamics and renormalization group calculations of dynamical
  critical properties},\ }\href {https://doi.org/10.1007/BF01316547} {\bibfield
   {journal} {\bibinfo  {journal} {Z Physik B}\ }\textbf {\bibinfo {volume}
  {23}},\ \bibinfo {pages} {377–380} (\bibinfo {year} {1976})}\BibitemShut
  {NoStop}%
\bibitem [{\citenamefont {{D}e {D}ominicis}(1976)}]{Dominicis1976}%
  \BibitemOpen
  \bibfield  {author} {\bibinfo {author} {\bibfnamefont {C.}~\bibnamefont {{D}e
  {D}ominicis}},\ }\bibfield  {title} {\bibinfo {title} {Techniques de
  renormalisation de la théorie des champs et dynamique des phénomènes
  critiques},\ }\href {https://doi.org/10.1051/jphyscol:1976138} {\bibfield
  {journal} {\bibinfo  {journal} {J. Phys. Colloques}\ }\textbf {\bibinfo
  {volume} {37}},\ \bibinfo {pages} {C1} (\bibinfo {year} {1976})}\BibitemShut
  {NoStop}%
\bibitem [{\citenamefont {Kloss}\ \emph {et~al.}(2012)\citenamefont {Kloss},
  \citenamefont {Canet},\ and\ \citenamefont {Wschebor}}]{Kloss2012}%
  \BibitemOpen
  \bibfield  {author} {\bibinfo {author} {\bibfnamefont {T.}~\bibnamefont
  {Kloss}}, \bibinfo {author} {\bibfnamefont {L.}~\bibnamefont {Canet}},\ and\
  \bibinfo {author} {\bibfnamefont {N.}~\bibnamefont {Wschebor}},\ }\bibfield
  {title} {\bibinfo {title} {Nonperturbative renormalization group for the
  stationary {Kardar-Parisi-Zhang} equation: Scaling functions and amplitude
  ratios in 1+1, 2+1, and 3+1 dimensions},\ }\href
  {https://doi.org/10.1103/PhysRevE.86.051124} {\bibfield  {journal} {\bibinfo
  {journal} {Phys. Rev. E}\ }\textbf {\bibinfo {volume} {86}},\ \bibinfo
  {pages} {051124} (\bibinfo {year} {2012})}\BibitemShut {NoStop}%
\bibitem [{\citenamefont {Benitez}\ \emph {et~al.}(2009)\citenamefont
  {Benitez}, \citenamefont {Blaizot}, \citenamefont {Chat\'e}, \citenamefont
  {Delamotte}, \citenamefont {M\'endez-Galain},\ and\ \citenamefont
  {Wschebor}}]{Benitez2009}%
  \BibitemOpen
  \bibfield  {author} {\bibinfo {author} {\bibfnamefont {F.}~\bibnamefont
  {Benitez}}, \bibinfo {author} {\bibfnamefont {J.-P.}\ \bibnamefont
  {Blaizot}}, \bibinfo {author} {\bibfnamefont {H.}~\bibnamefont {Chat\'e}},
  \bibinfo {author} {\bibfnamefont {B.}~\bibnamefont {Delamotte}}, \bibinfo
  {author} {\bibfnamefont {R.}~\bibnamefont {M\'endez-Galain}},\ and\ \bibinfo
  {author} {\bibfnamefont {N.}~\bibnamefont {Wschebor}},\ }\bibfield  {title}
  {\bibinfo {title} {Solutions of renormalization-group flow equations with
  full momentum dependence},\ }\href
  {https://doi.org/10.1103/PhysRevE.80.030103} {\bibfield  {journal} {\bibinfo
  {journal} {Phys. Rev. E}\ }\textbf {\bibinfo {volume} {80}},\ \bibinfo
  {pages} {030103} (\bibinfo {year} {2009})}\BibitemShut {NoStop}%
\bibitem [{\citenamefont {Mathey}\ \emph {et~al.}(2017)\citenamefont {Mathey},
  \citenamefont {Agoritsas}, \citenamefont {Kloss}, \citenamefont {Lecomte},\
  and\ \citenamefont {Canet}}]{Mathey2017}%
  \BibitemOpen
  \bibfield  {author} {\bibinfo {author} {\bibfnamefont {S.}~\bibnamefont
  {Mathey}}, \bibinfo {author} {\bibfnamefont {E.}~\bibnamefont {Agoritsas}},
  \bibinfo {author} {\bibfnamefont {T.}~\bibnamefont {Kloss}}, \bibinfo
  {author} {\bibfnamefont {V.}~\bibnamefont {Lecomte}},\ and\ \bibinfo {author}
  {\bibfnamefont {L.}~\bibnamefont {Canet}},\ }\bibfield  {title} {\bibinfo
  {title} {{K}ardar-{P}arisi-{Z}hang equation with short-range correlated
  noise: Emergent symmetries and nonuniversal observables},\ }\href
  {https://doi.org/10.1103/PhysRevE.95.032117} {\bibfield  {journal} {\bibinfo
  {journal} {Phys. Rev. E}\ }\textbf {\bibinfo {volume} {95}},\ \bibinfo
  {pages} {032117} (\bibinfo {year} {2017})}\BibitemShut {NoStop}%
\bibitem [{Note1()}]{Note1}%
  \BibitemOpen
  \bibinfo {note} {We have checked that the more common Wetterich regulator
  $\protect \hat r(\protect \hat p^2) = c / (\exp (\protect \hat p^2)-1)$
  yields the same behavior.}\BibitemShut {Stop}%
\bibitem [{\citenamefont {Canuto}\ \emph {et~al.}(2007)\citenamefont {Canuto},
  \citenamefont {Hussaini}, \citenamefont {Quarteroni},\ and\ \citenamefont
  {Zang}}]{2007-canuto--zang}%
  \BibitemOpen
  \bibfield  {author} {\bibinfo {author} {\bibfnamefont {C.}~\bibnamefont
  {Canuto}}, \bibinfo {author} {\bibfnamefont {M.~Y.}\ \bibnamefont
  {Hussaini}}, \bibinfo {author} {\bibfnamefont {A.}~\bibnamefont
  {Quarteroni}},\ and\ \bibinfo {author} {\bibfnamefont {T.~A.}\ \bibnamefont
  {Zang}},\ }\href {https://doi.org/10.1007/978-3-540-30728-0} {\emph {\bibinfo
  {title} {Spectral Methods: Evolution to Complex Geometries and Applications
  to Fluid Dynamics}}},\ Scientific Computation\ (\bibinfo  {publisher}
  {Springer},\ \bibinfo {address} {Berlin, Germany},\ \bibinfo {year}
  {2007})\BibitemShut {NoStop}%
\bibitem [{\citenamefont {Cox}\ and\ \citenamefont
  {Matthews}(2002)}]{2002-cox-matthews}%
  \BibitemOpen
  \bibfield  {author} {\bibinfo {author} {\bibfnamefont {S.}~\bibnamefont
  {Cox}}\ and\ \bibinfo {author} {\bibfnamefont {P.}~\bibnamefont {Matthews}},\
  }\bibfield  {title} {\bibinfo {title} {Exponential time differencing for
  stiff systems},\ }\href
  {https://doi.org/https://doi.org/10.1006/jcph.2002.6995} {\bibfield
  {journal} {\bibinfo  {journal} {Journal of Computational Physics}\ }\textbf
  {\bibinfo {volume} {176}},\ \bibinfo {pages} {430} (\bibinfo {year}
  {2002})}\BibitemShut {NoStop}%
\bibitem [{\citenamefont {Kassam}\ and\ \citenamefont
  {Trefethen}(2005)}]{2005-kassam-trefethen}%
  \BibitemOpen
  \bibfield  {author} {\bibinfo {author} {\bibfnamefont {A.-K.}\ \bibnamefont
  {Kassam}}\ and\ \bibinfo {author} {\bibfnamefont {L.~N.}\ \bibnamefont
  {Trefethen}},\ }\bibfield  {title} {\bibinfo {title} {Fourth-order
  time-stepping for stiff {PDEs}},\ }\href
  {https://doi.org/10.1137/S1064827502410633} {\bibfield  {journal} {\bibinfo
  {journal} {SIAM Journal on Scientific Computing}\ }\textbf {\bibinfo {volume}
  {26}},\ \bibinfo {pages} {1214} (\bibinfo {year} {2005})}\BibitemShut
  {NoStop}%
\bibitem [{\citenamefont {Pr\"ahofer}\ and\ \citenamefont
  {Spohn}(2004)}]{Praehofer04}%
  \BibitemOpen
  \bibfield  {author} {\bibinfo {author} {\bibfnamefont {M.}~\bibnamefont
  {Pr\"ahofer}}\ and\ \bibinfo {author} {\bibfnamefont {H.}~\bibnamefont
  {Spohn}},\ }\bibfield  {title} {\bibinfo {title} {Exact scaling functions for
  one-dimensional stationary {KPZ} growth},\ }\href
  {https://doi.org/10.1023/B:JOSS.0000019810.21828.fc} {\bibfield  {journal}
  {\bibinfo  {journal} {J. Stat. Phys.}\ }\textbf {\bibinfo {volume} {115}},\
  \bibinfo {pages} {255} (\bibinfo {year} {2004})}\BibitemShut {NoStop}%
\bibitem [{\citenamefont {Tarpin}\ \emph {et~al.}(2018)\citenamefont {Tarpin},
  \citenamefont {Canet},\ and\ \citenamefont {Wschebor}}]{Tarpin2018}%
  \BibitemOpen
  \bibfield  {author} {\bibinfo {author} {\bibfnamefont {M.}~\bibnamefont
  {Tarpin}}, \bibinfo {author} {\bibfnamefont {L.}~\bibnamefont {Canet}},\ and\
  \bibinfo {author} {\bibfnamefont {N.}~\bibnamefont {Wschebor}},\ }\bibfield
  {title} {\bibinfo {title} {Breaking of scale invariance in the time
  dependence of correlation functions in isotropic and homogeneous
  turbulence},\ }\href {https://doi.org/10.1063/1.5020022} {\bibfield
  {journal} {\bibinfo  {journal} {Physics of Fluids}\ }\textbf {\bibinfo
  {volume} {30}},\ \bibinfo {pages} {055102} (\bibinfo {year}
  {2018})}\BibitemShut {NoStop}%
\bibitem [{\citenamefont {Gorbunova}\ \emph {et~al.}(2021)\citenamefont
  {Gorbunova}, \citenamefont {Pagani}, \citenamefont {Balarac}, \citenamefont
  {Canet},\ and\ \citenamefont {Rossetto}}]{Gorbunova2021scalar}%
  \BibitemOpen
  \bibfield  {author} {\bibinfo {author} {\bibfnamefont {A.}~\bibnamefont
  {Gorbunova}}, \bibinfo {author} {\bibfnamefont {C.}~\bibnamefont {Pagani}},
  \bibinfo {author} {\bibfnamefont {G.}~\bibnamefont {Balarac}}, \bibinfo
  {author} {\bibfnamefont {L.}~\bibnamefont {Canet}},\ and\ \bibinfo {author}
  {\bibfnamefont {V.}~\bibnamefont {Rossetto}},\ }\bibfield  {title} {\bibinfo
  {title} {Eulerian spatiotemporal correlations in passive scalar turbulence},\
  }\href {https://doi.org/10.1103/PhysRevFluids.6.124606} {\bibfield  {journal}
  {\bibinfo  {journal} {Phys. Rev. Fluids}\ }\textbf {\bibinfo {volume} {6}},\
  \bibinfo {pages} {124606} (\bibinfo {year} {2021})}\BibitemShut {NoStop}%
\bibitem [{\citenamefont {Gosteva}\ \emph
  {et~al.}(2025{\natexlab{b}})\citenamefont {Gosteva}, \citenamefont
  {Brachet},\ and\ \citenamefont {Canet}}]{Gosteva2025Euler}%
  \BibitemOpen
  \bibfield  {author} {\bibinfo {author} {\bibfnamefont {L.}~\bibnamefont
  {Gosteva}}, \bibinfo {author} {\bibfnamefont {M.}~\bibnamefont {Brachet}},\
  and\ \bibinfo {author} {\bibfnamefont {L.}~\bibnamefont {Canet}},\ }\bibfield
   {title} {\bibinfo {title} {Emergent dynamical scaling in the inviscid limit
  of {3D} stochastic {N}avier-{S}tokes equation with thermal noise},\
  }\href@noop {} {\bibfield  {journal} {\bibinfo  {journal} {arXiv:2507.05811}\
  } (\bibinfo {year} {2025}{\natexlab{b}})}\BibitemShut {NoStop}%
\end{thebibliography}

%

\end{document}